\documentclass[12pt]{iopart}

\newtheorem{observation}{Observation}

\newtheorem{definition}{Definition}
\newtheorem{theorem}{Theorem}
\newtheorem{lemma}{Lemma}

\begin{document}
\title{Nonclassicality witnesses and entanglement creation}

\author{F. E. S. Steinhoff} 

\address{Instituto de F\'\i sica ``Gleb Wataghin'', Universidade Estadual de Campinas, 13083-970, Campinas, SP, Brazil}

\ead{steinhof@ifi.unicamp.br}

\pacs{03.67.Mn, 03.65.Ud, 42.50.Dv}
\date{\today}

\begin{abstract}
Several definitions of classicality are considered, such as P-representability, generalized coherent states and separable states. These notions are treated under a simple and general definition based on convex sets, which enables the use of the Hahn-Banach theorem to separate classical states from nonclassical ones. Nonclassicality linear witnesses are constructed, based on the observables available in a given physical situation. Some examples of nonclassical states are considered, with detection schemes available nowadays. Reviewing the concept of entanglement potential from a different perspective, it is shown that in some contexts an arbitrary single-system nonclassical state can be converted into a bipartite entangled state, provided a generalized controlled-displacement (e.g., beam-splitter, CNOT gate) is available. Also, this entanglement can be detected in a simple way using the nonclassicality witness of the original single-system state. Finally, we extend the discussion to multipartite states, proposing alternative ways to generate and classify multipartite entanglement.   
\end{abstract}
\maketitle

\section{Introduction}

Characterization of quantum phenomena has a central importance both from a foundational level as well as from possible practical applications. Since Bell's Theorem \cite{bell}, physical events are taken as intrinsically random, while the classical world we experience is seen as a result of the law of large numbers \cite{vneumann} and the unavoidable effect of interactions with the environment \cite{zurek}. But even if one considers a local description of a physical system, nonclassical effects may appear, i.e., effects that have no counterpart in a classical description of the world. For example, a single-mode state with negative Glauber-Sudarshan function necessarilly has some kind of optical correlation \cite{mandel}; even a single spin-1 object can be proven to present nonclassical effects \cite{klyachko2}. Also, the potential applications of nonclassical states are surging in a fast pace. Entanglement alone can be used as a resource in many tasks that would be impossible with separable states \cite{nielsen, horodecki}. As we will try to show in the present work, in some physical contexts an arbitrary nonclassical state has a potential to generate some entangled state, hence being potentially usefull as a resource. 

However, besides its obvious relevance, nonclassicality is a term with many different definitions nowadays. To say that a state is nonclassical could mean that it has higher order correlations in interferometric experiments \cite{mandel}, or that it cannot be created solely via local operations and classical comunication \cite{werner}. It can mean that for some apparatus, its statistical quantities like transition probabilities or mean values do not follow usual probability predictions \cite{zurek}, or that it is not in some orbit of a dynamical symmetry group of a system of interest \cite{zhang, klyachko}, or even that it can outperform tasks made with a classical computer \cite{galvao}. All these different notions share one common structure: the states understood as classical form a closed convex set, while those taken as nonclassical are outside this set \footnote{We are excluding the nonclassicality associated to states with some degree of quantum discord, since classical states do not form a convex set; also, quantum discord only makes sense for composite systems.}. 

We take advantage of this feature in order to give a general method of detection of nonclassicality, irrespective of the meaning that is assembled to this word. This approach is based on standard procedures for detection of quantum entanglement \cite{toth, guhne, sperling2} and is practical in the sense that uses the observables relevant to the physical situation at hand. Focusing in single-system's nonclassicality, it is shown that any nonclassical state has the potential to generate entanglement, provided some controlled displacement operation is available. This notion of entanglement potential was originally proposed in \cite{asboth} for the physical context of quantum optics, but here we extend it and show whether it could have a more consistent definition. Practical examples of controlled displacements operations are the beam-splitter and CNOT gates. This method can be further extended to multipartite systems, in order to obtain an intuition about the different classes of multipartite entanglement that arise in quantum information theory. We will see how different notions of classicality that appear in the creation of a multipartite state affect its classification.

The manuscript is organized as follows. In Section II some usual classicality definitions are considered and then are treated under a general definition, which turns the set of classical states into a closed convex set. In Section III, the existence of observables known as witnesses detecting the nonclassicality of a state is shown using a corollary of the Hahn-Banach theorem. A method to construct nonclassicality witnesses using the observables available in a given physical situation is developed. Also, some examples of states detectable by this method are shown, with significant simplicity in comparison with standard procedures such as calculation of the $P$-function. In Section IV we redefine the concept of entanglement potential in a more consistent fashion, in order to show that any single-system nonclassical state has the potential to be converted into a bipartite entangled state. This conversion is done by an unitary operation which is a generalized version of a CNOT gate for a two-qubit system. Also, the nonclassicality witness of the original single-system state can be used to detect the entanglement of the output state, a simplification with many potential applications. In Section V, the entanglement potential approach is extended to multipartite states. The usual $W$ and $GHZ$ classes of three pure qubits are constructed taking different notions of classicality into account, and then an intuitive procedure to classify the different classes of entanglement is proposed, based on Jordan normal forms. The four inequivalent classes of tripartite entanglement for two qutrits and a qubit \cite{cornelio} are also constructed. In Section VI we consider some further details and give the concluding remarks.

\section{Convex definitions of classicality}

Let us introduce some concepts and terminology. Hilbert spaces will be denoted by $\mathcal{H}$, with $d_{\mathcal{H}}$ being its dimension. Operators will be indicated with a $\hat{}$ symbol, with exception of the identity operator, which will be simply denoted by $I$. In some situations we will need the set of bounded operators of some topological vector space $\mathcal{S}$; this set will be denoted by $B(\mathcal{S})$. 

\subsection{Canonical coherent states}

The usual example of a quantum state that is taken as classical or quasi-classical is a coherent state as it appears in the theory of quantum optics. A state in this class will be refereed as a Canonical Coherent State (CCS) in what follows. For a review about the subject, \cite{mandel, zhang}. The infinite-dimensional Hilbert space $\mathcal{H}$ is spanned by the discrete orthonormal Fock basis $\{|n\rangle\}_{n=0}^{\infty}$, with $n$ a natural number. The annihilation and creation operators in Fock basis are respectively given by
\begin{eqnarray}
\hat{a}|n\rangle=\sqrt{n}|n-1\rangle ; \ \ \hat{a}^{\dagger}|n\rangle = \sqrt{n+1}|n+1\rangle
\end{eqnarray}
satisfying $[\hat{a},\hat{a}^{\dagger}]=\hat{1}$. A coherent state $|\alpha\rangle$ is by definition an eigenstate of the annihilation operator $\hat{a}$ such that $\hat{a}|\alpha\rangle=\alpha|\alpha\rangle$. Since measurements in quantum optics are usually done through absorption of photons, coherent states are invariant under such operations. Also, these states minimize the uncertainty relation associated with measurements of position and momentum, being thus the states closest to classical behaviour. CCSs can also be expressed as displacements of the vaccum state $|0\rangle$ in a phase space with classical structure:
\begin{eqnarray}
|\alpha\rangle = \hat{D}(\alpha)|0\rangle
\end{eqnarray}
where $\hat{D}(\alpha)=e^{\alpha \hat{a}^{\dagger}-\alpha^*\hat{a}}$ is called the displacement operator. Displacement operators form an unitary representation of the Heisenberg-Weyl (HW) group under the multiplication rule $\hat{D}(\alpha)\hat{D}(\beta)=e^{(\alpha\beta^*-\alpha^*\beta)/2}\hat{D}(\alpha+\beta)$. Thus, CCSs are the HW-orbit of vaccum and this property will be used to construct generalized version of coherent states. Also, the set of coherent states is an overcomplete basis for $\mathcal{H}$.

In general, a state $\hat{\rho}$ can be expressed uniquelly in terms of its Glauber-Sudarshan representation:
\begin{eqnarray}
\hat{\rho} = \int_{-\infty}^{+\infty} P(\alpha)|\alpha\rangle\langle\alpha |d^2\alpha  
\end{eqnarray}
where
\begin{eqnarray}
d^2\alpha = d\Re(\alpha)d\Im(\alpha); \ \ \ \frac{1}{\pi}\int_{-\infty}^{+\infty}|\alpha\rangle\langle\alpha|d^2\alpha=I \label{povm}
\end{eqnarray}
and $P(\alpha)$ is its Glauber-Sudarshan function, satisfying 
\begin{eqnarray}
 \int_{-\infty}^{+\infty} P(\alpha)d^2\alpha=1. 
\end{eqnarray}
A state is then called classical or P-representable if its Glauber-Sudarshan function is nonnegative and less singular than the Dirac delta distribution. In this situation, $P(\alpha)$ is a true probability density distribution, with the set of operators $\pi^{-1}|\alpha\rangle\langle\alpha|$ forming a Positive-Operator Valued Measure (POVM) under (\ref{povm}) \cite{peres}. It is straightforward that classical states form a closed convex set, with pure coherent states as their extremal points. In this sense, a state is nonclassical if it is not P-representable. Indeed, in this situation the nonclassicality of the state can be observed via optical interferometric experiments \cite{mandel}, which reveal quantum correlations of a certain degree.

\subsection{Generalized coherent states}

Various attempts to generalize the concept of coherent states for arbitrary physical systems were proposed \cite{zhang}, extending some property of the CCS that was relevant for the problem at hand. Perhaps the most sucessfull approach is that of Generalized Coherent States, developed independently by Perelomov \cite{perelomov} and Gillmore \cite{gillmore}. The brief and informal treatment given here follows the pioneering works of Klyachko \cite{klyachko} and followers \cite{viola}, which give a general connection between this notion of classicality and entanglement theory. 

Given a Lie group $\mathsf{G}$, a pure Generalized Coherent State (GCS) $|\alpha_{\mathsf{G}}\rangle$ associated to this group is an element of the orbit of $\mathsf{G}$ in a reference state, choosen by physical/symmetry reasons. For example, we have CCSs as the orbit of the Heisenberg-Weyl group on vaccum, which is the ground state of the harmonic oscillator hamiltonian and thus is a maximally degenerate and symmetric state. In many cases, the observables available in some physical context have the structure of a Lie-algebra, given by its commutation relations. To the relevant Lie-algebra $\mathsf{g}$ one associates a Lie-group $\mathsf{G}$ via the exponential mapping. For example, if $\mathsf{g}$ is semisimple, we can take a highest weight vector as a reference state; the $\mathsf{G}$-orbit of this state will be the set of generalized coherent states. A state will be then classical if it can be written as 
\begin{eqnarray}
\hat{\rho} = \int_{-\infty}^{+\infty}P_{\mathsf{G}}(\alpha)|\alpha_{\mathsf{G}}\rangle\langle\alpha_{\mathsf{G}}|d\mu(\alpha)
\end{eqnarray}  
for some probability density function $P_{\mathsf{G}}$ and some probability measure $d\mu$ (e.g., the Haar-measure of the group $G$). The set of GCS is also an overcomplete basis for $\mathcal{H}$. This notion of classicality is particularly usefull when the physical system is subject to some superselection rule. In this situation, some observables are forbidden and the allowed measurements are restricted to a smaller set of observables. A basic example occurs when we analise angular momentum of a system of particles. The allowed measurements are those of total angular momentum $\mathbf{J}=\sum_k \mathbf{J}_k$, with $\mathbf{J}_k$ being the angular momentum operator of the kth particle. Since the components of $\mathbf{J}$ are the generators of $\mathsf{su}(2)$ Lie-algebra, we are restricted to the observables of this algebra and the corresponding $\mathsf{SU}(2)$ Lie-group of unitary operations will correspond to the symmetries of the system. 

\subsection{Product states}

For a composite system $\mathcal{H}=\bigotimes_i \mathcal{H}_i$, we can define the classicality associated with product states $|\phi\rangle = \bigotimes_j|\phi_j\rangle$: a state will be classical or separable if it can be approximated in trace norm by convex combinations of product states \cite{werner, horodecki}:
\begin{eqnarray} 
\hat{\rho}_c = \sum_i p_i\bigotimes_j\hat{\rho}^{(i)}_j \label{separable}
\end{eqnarray}
where $\sum_i p_i =1$, $p_i\geq 0$ and $\hat{\rho}^{(i)}_j\in B(\mathcal{H}_j)$ . This is of course equivalent to say that a state is separable if it is a convex combination of pure product states. Thus, pure product states are extremal points of the convex set of classical states and form a basis for the global Hilbert space $\mathcal{H}$. States that cannot be decomposed as (\ref{separable}) will be then called entangled. Equivalently, separable states are the most general states that can be generated under the restriction to Stochastic Local Operations and Classical Communication (SLOCC). 

\subsection{General definition}

By the examples considered, we define a classical basis as a fixed basis $C=\{|c_{\nu}\rangle\}$ of a Hilbert space $\mathcal{H}$, where the index $\nu$ can assume discrete or continuous values, depending on $\mathcal{H}$. Being a basis of a Hilbert space, $C$ will at least satisfy some POVM relation $\int |c_i\rangle\langle c_i|d\mu=I$ for some convenient probability measure $\mu$, so that we can expand an arbitrary state in terms of the elements of $C$. The reasons to choose a specific basis will depend on the problem at hand. This basis can be, for example, the pointer basis associated to the measurements performed on a system \cite{zurek}; or it can be choosen due to its inability to outperform tasks done in classical computers \cite{galvao}. More important is the notion of a classical state:
\begin{definition}
A state $\hat{\rho}$ is classical if it can be approximated, in some operator norm \footnote{The usual norm used by physicists is the trace norm, given that $Tr(M\rho)$ is the mean value of observable $M$ if the state of the system is $\hat{\rho}$. Here we allow any operator norm, since classicality can be understood as a more abstract notion in many situations.}, by states of the form
\begin{eqnarray}
\hat{\rho}_c = \sum_i p_i|c_i\rangle\langle c_i| \label{cstates}
\end{eqnarray}
with $p_i\geq 0$ and $\sum_i p_i=1$, i.e., convex combinations of pure classical states. Equivalently, $\hat{\rho}$ is classical if there exists a probability measure $\mu$ and a probability density distribution $P$ such that
\begin{eqnarray*}
\hat{\rho}=\int_{\Omega}P(c)|c\rangle\langle c|d\mu(c) 
\end{eqnarray*}
with $\Omega$ being the probability space considered in the physical context.
\end{definition}
Thus, classical states form a closed convex set, a very useful property for their characterization. Depending on the classical basis at hand, there will be states that cannot be approximated by states of the form (\ref{cstates}) and are not contained in the set of classical states. Those states will be called nonclassical. 

Another usefull concept is of a classical operation \cite{sperling}. A operation $\Lambda$  is called classical if for any classical state $\rho_c$, we have $\Lambda(\rho_c)/Tr[\Lambda(\rho_c)]$ is a classical state; otherwise it will be called a nonclassical operation. It is easilly seen that the set of classical operations is a monoid, i.e., a semigroup with identity. We will be particularly concerned with the effect of nonclassical operations on the elements of $C$.

\section{Nonclassicality witnesses}

The theorems given here are adaptations of results about entanglement witnesses from references \cite{horodecki2, terhal, toth, sperling2}; the proofs are shortened versions of the demonstrations of these references, which should be consulted in case more details are needed.

\subsection{Linear witnesses and the Hahn-Banach Theorem}

Since the set of classical states is convex and closed, intuitively one can separate it from nonclassical states using suitable hyperplanes. To comprise the situation of infinite-dimensional systems we use the following corollary of the Hahn-Banach Theorem \cite{edwards}:
\begin{theorem}
Given two closed convex sets $S_1$ and $S_2$ in a real Banach space, one of which is compact, then there exists a bounded linear functional $f$ that separates the two sets, i.e, there exists $\xi\in\Re$ such that $f(s_1)<\xi\leq f(s_2)$ for all $s_1\in S_1$, $s_2\in S_2$. 
\end{theorem}
The value of the functional $f$ on an arbitrary state can be represented as $f(\hat{\rho})=Tr(\hat{W}\hat{\rho})$, with $\hat{W}\in B(\mathcal{H})$, since the dual space of trace-class operators of the Banach space is isomorphic to the set of bounded operators. So, we have
\begin{lemma}
For every nonclassical state $\hat{\rho}$ there exists an operator $\hat{W}$ such that $Tr(\hat{W}\hat{\rho})<0$ and $Tr(\hat{W}\hat{\sigma})\geq 0$, for all classical $\hat{\sigma}$.
\end{lemma}
\textbf{Proof:} From Theorem 1 and observations following it, we have that there exists a bounded hermitean operator $\tilde{W}$ and $\xi\in\Re$ such that
\begin{eqnarray}
Tr(\tilde{W}\hat{\rho})<\xi\leq Tr(\tilde{W}\hat{\sigma})
\end{eqnarray}
for all classical state $\hat{\sigma}$. Since $\xi=Tr(\xi I)$ and defining $\hat{W}=\tilde{W}-\xi I$, the assertion follows, QED.

We are led naturally to the following definition, using the same terminology of \cite{terhal}: 
\begin{definition}
A bounded hermitean operator $\hat{W}$ is a nonclassicality witness if $Tr(\hat{W}\hat{\sigma})\geq 0$, for any classical state $\hat{\sigma}$ and there exists at least one state $\hat{\rho}$ such that $Tr(\hat{W}\hat{\rho})<0$.
\end{definition}
An immediate implication of Lemma 1 is 
\begin{theorem}
A state $\hat{\sigma}$ is classical iff $Tr(\hat{W}\hat{\sigma})\geq 0$ for all nonclassicality witnesses $\hat{W}$.
\end{theorem}
Thus, for any nonclassical state $\hat{\rho}$ there is an observable $\hat{W}$ which detects its nonclassicality, i.e., such that the mean value of $\hat{W}$ on this state is negative. Geometrically, the states for which $\langle W\rangle=Tr(\hat{W}\hat{\rho})=0$ define a hyperplane which divides the convex set of classical states - obeying $\langle \hat{W}\rangle\geq 0$ - from the compact set defined by the point $\hat{\rho}$ - obeying $\langle \hat{W}\rangle <0$. It is easy to see that given a nonclassicality witness $\hat{W}$ some nonclassical states will be not detected by $\hat{W}$: to fully determine the set of classical states in general one needs an infinite number of witnesses, as is also clear from Theorem 2.

\subsection{Construction of witnesses}

Let us now introduce the notion of optimization of entanglement witnesses. We say that a witness $\hat{W}_1$ is finer than another witness $\hat{W}_2$ when $\hat{W}_1$ detects all nonclassical states detected by $\hat{W}_2$ and also some more. We call a witness $\hat{W}_{opt}$ optimal when no other witness is finer than $\hat{W}_{opt}$. Theorem 2 can be thus simplified, since optimal entanglement witnesses are enough to characterize the set of classical states. Borrowing some standard procedures in quantum information theory \cite{toth, guhne, sperling2}, we develop now a method to construct nonclassicality witnesses from observables available in practice. Given an observable $\hat{M}$ and a classical basis $\hat{C}$, let us define the hermitean operator
\begin{eqnarray}
\hat{W}_M = \lambda(\hat{M}) I - \hat{M} \label{operator}
\end{eqnarray}
where $\lambda(\hat{M}) =max_{|c\rangle}\{\langle c|\hat{M}|c\rangle:|c\rangle \in C\}$, that is, the maximization is done over the set of classical. 
\begin{observation}
Observable (\ref{operator}) has positive mean value for any classical state.
\end{observation}
\textbf{Proof:} An arbitrary classical state, according to Definition 1, is given by
\begin{eqnarray}
\hat{\rho}=\int_{\Omega}P(c)|c\rangle\langle c|d\mu(c)
\end{eqnarray}
for some probability space $\Omega$, probability measure $\mu$ and probability distribution $P$. Taking the mean value of $\hat{M}$ on this state and using the linearity of the trace, we have:
\begin{eqnarray}
Tr(\hat{M}\hat{\rho}) &=& \int_{\Omega}P(c)Tr(\hat{M}|c\rangle\langle c|)d\mu(c) \\
&=& \int_{\Omega}P(c)\underbrace{\langle c|\hat{M}|c\rangle}_{\leq\lambda(\hat{M})} d\mu(c) \leq \lambda(\hat{M})
\end{eqnarray}
thus implying $Tr(\hat{W}_M\rho)=\lambda(\hat{M})-Tr(\hat{M}\rho)\geq 0$, QED.

So, if the mean value of $\hat{W}$ is negative for some state, this state is nonclassical and we say that its nonclassicality is detected by the operator $\hat{W}$. It is obvious that (\ref{operator}) is optimal in this sense. That there exists any state detectable by this method will be shown in the examples of the next section, but one could consider the special case of entanglement witnesses, where the classical basis is constituted of pure product states. 

The formulation in terms of (\ref{operator}) enables a practical way of detection of nonclassicality in terms of available measurements, since the detection is reduced to verifying wheter $Tr(\hat{W}_M)<0$, i.e., $\lambda(\hat{M}) < Tr(\hat{M}\rho)$; in words, an entangled state is nonclassical if its mean value for some observable $\hat{M}$ exceeds the maximal mean value of $\hat{M}$ on the set of classical states. The value $\lambda(\hat{M})$ can be very difficult to calculate exactly in some cases. It is easily seen, however, that any real value $\lambda \geq \lambda(\hat{M})$ results in an operator $\hat{W}=\lambda I - \hat{M}$ that has positive mean value on classical states as well and could be used as a witness. This witness will be of course non-optimal, but perhaps easier to implement in practice. Also, as stated in \cite{sperling2}, an arbitrary optimal witness can always be expressed as (\ref{operator}), for some suitable observable $\hat{M}$. In this manuscript we are mostly concerned with what can be done in practice with the observables available to the experimentalist. For example, we see that some measurements can be discarded. If we have, for example, an observable that is diagonal in classical basis, then it is straightforward that (\ref{operator}) cannot detect any nonclassicality. Indeed, any $\hat{W}_M$ that is positive semidefinite cannot detect nonclassicality and an observable diagonal in a classical basis trivially results in a positive semidefinite $\hat{W}_M$.

\subsection{Applications}

\subsubsection{Coherent states}

The standard examples of nonclassical states in this case are Fock states, cat states and squeezed states. Witnesses for Fock states are particularly useful, since it is customary to express an arbitrary density matrix in Fock basis, $\hat{\rho}=\sum_{i,j}\rho_{ij}|i\rangle\langle j|$. To detect the nonclassicality of a single-mode Fock state, we define $\hat{M}_n=|n\rangle\langle n|$ and calculate 
\begin{eqnarray}
\lambda(\hat{M}_n)= \frac{e^{-n}n^n}{n!}
\end{eqnarray}
So, we have $\lambda(\hat{M}_n)<Tr(\hat{M}_n|n\rangle\langle n|)=1$ and thus any number state is nonclassical. This method is much simpler than the usual procedure of calculating the Glauber-Sudarshan $P$-function. Also, this results imposes a threshold on the photocount distribution of an arbitrary state $\hat{\rho}$, $p_n =\langle n|\hat{\rho}|n\rangle=Tr(\hat{M}_n\rho)$, which when violated, implies nonclassicality. Experimentally, there are many methods to obtain $p_n$ \cite{mandel}. As an example, the photocount distribution of a Single-Photon Added Thermal State (SPATS) is given by \cite{bellini}
\begin{eqnarray}
p_n= \frac{1}{\bar{n}(\bar{n}+1)}\left(\frac{\bar{n}}{\bar{n}+1}\right)^n n
\end{eqnarray}
where $\bar{n}$ represents the mean photonumber of the state. It is very easy to find values for $\bar{n}$ such that $p_n>\lambda(\hat{M}_n)$. For $n=1$, we have the condition $(\bar{n}+1)<e$; this could be measured in a single photocounting module, for example. The following vaccum-squeezed state 
\begin{eqnarray}
|\phi\rangle = \sqrt{1-q^2}\sum_{k=0}^{\infty}q^n|2n\rangle; \ \ 0<q<1
\end{eqnarray}
can be proven to be nonclassical in a similar fashion. Defining $\hat{M}_{\phi}=|\phi\rangle\langle\phi|$, it is straightforward that $\lambda(\hat{M}_{\phi})=1-q^2$, implying $\lambda(\hat{M}_{\phi})<Tr(\hat{M}_{\phi}|\phi\rangle\langle\phi|)=1$ and the state is nonclassical. Finally, given a coherent state $|\alpha\rangle$ ($\alpha\neq 0$), a cat-state 
\begin{eqnarray}
|\eta\rangle = \frac{1}{\sqrt{2}}(|\alpha\rangle +|-\alpha\rangle) 
\end{eqnarray}
is trivially shown to be nonclassical by this method. Defining $\hat{M}_{\eta}=|\eta\rangle\langle\eta|$, it is trivial that $\lambda(\hat{M}_{\eta})<1=Tr(\hat{M}_{\eta}|\eta\rangle\langle\eta$. In comparison with the usual method of finding the Glauber-Sudarshan function, our method is clearly simpler - at least in the examples considered.   

\subsection{$SU(2)$ coherent states}

When the classical basis is consitituted of $SU(2)$ coherent states, one is restricted to total angular momentum observables $J_n\equiv\mathbf{J}\cdot\mathbf{n}$. However, a single observable $J_n$ for some direction $\mathbf{n}$ will not be usefull to construct a nonclassicality witness. Since generalized coherent states are maximum weight vectors for some $J_n$, the maximal value of $\langle\alpha_G|J_n|\alpha_G\rangle$ will be the maximal mean value of $J_n$ and the operator $W=\lambda(J_n)I-J_n$ will be positive semidefinite. So we must consider the mean values of higher order terms, such as $(J_x)^2$, $J_xJ_y$, $J_xJ_z$, $(J_y)^2$, \ldots, in order to construct nonclassicality witnesses. Although it can be hard to design apparatuses to implement these higher order observables, one can obtain their mean values (multipolar moments) as described in \cite{newton}. The higher order moments are related to the moments of $J_n$ in different directions $\mathbf{n}$ through a system of linear equations. The price paid for such simplification is that the system of equations is generally redundant and the number of different directions to be measured can grow quickly with increasing number of particles. When dealing with Stern-Gerlach apparatuses, a perhaps easier alternative is to consider Feynman filters \cite{gale}. The idea is that an arbitrary rank-one projector is the eigenstate of some $J_n$. The examples of observables considered here are all rank-one projectors, so one would need simply to find the proper direction $\mathbf{n}$ and record the rate at which the detector in this direction is hit; rates at other directions should be discarded, or used to infer mean values of higher-rank observables. 

The expression for generic $\mathsf{SU}(2)$ coherent states \cite{zyc} is
\begin{eqnarray}
|z\rangle = \left(\frac{1}{1+|z|^2}\right)^{j}\sum_{m=-j}^{m=j}z^{j+m}\sqrt{{2j\choose j+m}}|j,m\rangle
\end{eqnarray}
where $|j,m\rangle$ are the simultaneous eigenstates of $J_z$ and $J^2$, also known as Dicke states. As a simple example, we consider a spin-$1$ system, with
\begin{eqnarray*}
J_x = \frac{1}{\sqrt{2}}\left(\begin{array}{c c c}
{0}&{1}&{0}\\
{1}&{0}&{1}\\
{0}&{1}&{0}
\end{array}\right); \ \ J_y = \frac{1}{\sqrt{2}}\left(\begin{array}{c c c}
{0}&{-i}&{0}\\
{i}&{0}&{-i}\\
{0}&{i}&{0}
\end{array}\right); \ \ J_z = \left(\begin{array}{c c c}
{-1}&{0}&{0}\\
{0}&{0}&{0}\\
{0}&{0}&{1}
\end{array}\right)
\end{eqnarray*}
Coherent states in this case are given by
\begin{eqnarray}
|z\rangle = \left(\frac{1}{1+|z|^2}\right)\left(|-1\rangle+z\sqrt{2}|0\rangle +z^2|1\rangle\right)
\end{eqnarray}
We consider the quadrupole operator
\begin{eqnarray} 
M &=&(J_x)^2-(J_y)^2=\frac{1}{2}\left(\begin{array}{c c c}
{0}&{0}&{1} \\
{0}&{0}&{0} \\
{1}&{0}&{0}
\end{array}\right) 
\end{eqnarray}
whose eigenvectors are $|\psi_{\pm}\rangle=(1/\sqrt{2})|-1\rangle\pm |+1\rangle$. The mean value of $M$ in sn srbitrary coherent state is 
\begin{eqnarray}
\langle z|M|z\rangle = \frac{2\Re (z^2)}{(1+|z|^2)^2}=\frac{2(z^2_r-z^2_i)}{(1+z^2_r+z^2_i)^2}
\end{eqnarray}
where $z=z_r+iz_i$. The maximal is clearly $\lambda(M)=1/2$ and we conclude that $|\psi_+\rangle$ is nonclassical, given that $Tr(M|\psi_+\rangle\langle\psi_+|)=1$. Repeating the reasoning for the observable $-M$, we conclude that $|\psi_-\rangle$ is nonclassical as well. We stress that this was determined through measurement of fewer observables than the ones needed to perform full quantum tomography. The same reasoning can be applied to show that state $|0\rangle$ is nonclassical. Consider the following observable
\begin{eqnarray}
M'=I-(J_z)^2 = \left(\begin{array}{c c c}
{0}&{0}&{0}\\
{0}&{1}&{0}\\
{0}&{0}&{0}
\end{array}\right)=|0\rangle\langle 0|
\end{eqnarray}
Then it is easy to calculate $\lambda(M')=1/2$ and we have that $|0\rangle$ is nonclassical. 

\subsection{$d$-level systems}

For a system with orthonormal classical basis $\{|0\rangle, |1\rangle, \ldots, |d-1\rangle\}$, we can form pure nonclassical states through superpositions: 
\begin{eqnarray} 
|\psi\rangle = \sum_{i=0}^{d-1} c_i |i\rangle \label{superposition}
\end{eqnarray}
Using the observable $M_{\psi}=|\psi\rangle\langle\psi|$ we obtain $\lambda(M_{\psi})=max\{|c_i|^2\}$ and then we have $\lambda(M_{\psi})<1=Tr(M_{\psi}|\psi\rangle\langle\psi|)$, whenever $|\psi\rangle$ is the superposition of two or more classical states. As an example, one could consider the eigenvectors of $J_z$ as the classical basis: $C=\{|-s\rangle, |-s+1\rangle, \ldots, |s-1\rangle, |s\rangle\}$. Then, it is straightforward that the eigenvectors of $\mathbf{J}\cdot\mathbf{n}$ for any direction different from $z$ (or $-z$) are superpositions of the elements of $C$ and are thus nonclassical. Comparing with the case of $\mathsf{SU}(2)$ coherent states, we see that the physical system is the same, but the meaning of classicality for each case is radically different. 

For mixed states we will use further in the manuscript the notion of Superposition Number \cite{sperling}, analogously to Schmidt Number in entanglement theory. First, let us define the Superposition Number of a pure state $|\psi\rangle$. According to \cite{sperling}, this is the minimal number of pure classical states needed to express $|\psi\rangle$. But here, as we are defining pure classical states to form an orthonormal basis (complete set), there is no worry about possible ambiguities; the vector $|\psi\rangle$ is expressed uniquely as a linear combination of elements of $C$, as in (\ref{superposition}). Thus, the superposition number $r(\psi)$ of (\ref{superposition}) is the number of non null coefficients $c_i$. 

A mixed state $\rho$ will have Superposition Number $N_S(\rho)=k$ if: (a) for any decomposition of $\rho=\sum_i p_i|\psi_i\rangle\langle\psi_i|$ at least one of the vectors $|\psi_i\rangle$ has at least Superposition Number $k$; (b) there exists a decomposition of $\rho$ with all vectors $|\psi_i\rangle$ with Superposition Number at most $k$. We can rewrite this as the following optimization problem \cite{sperling}:
\begin{eqnarray}
N_S(\rho)=inf\{sup(r(\psi_i)):\rho=\sum_ip_i|\psi_i\rangle\langle\psi_i|\} \label{ns}
\end{eqnarray}
where the infimum is computed considering all decompositions $\rho=\sum_ip_i|\psi_i\rangle\langle\psi_i|$ and $sup(r(\psi_i))$ is the supremum value for a fixed decomposition. For classical states $N_S(\rho_C)=1$, so for $N_S(\rho)>1$ state $\rho$ is nonclassical. The analogy with the Schmidt number will be shown to be even stronger: for an orthonormal classical basis, one is able to convert the superposition number of a single state into the Schmidt number of a bipartite state  using suitable global interactions.

For a two-level system, nonclassicality can be completely characterized. In this case, the classical basis is given by $\{|0\rangle, |1\rangle\}$. Using the Pauli matrices,
\begin{eqnarray}
\sigma_x = \left(\begin{array}{c c}
{0}&{1}\\
{1}&{0}
\end{array}\right); \ \ 
\sigma_y = \left(\begin{array}{c c}
{0}&{-i}\\
{i}&{0}
\end{array}\right); \ \
\sigma_z = \left(\begin{array}{c c}
{1}&{0}\\
{0}&{-1}
\end{array}\right)
\end{eqnarray}
it is well known that an arbitrary density matrix can be expressed as
\begin{eqnarray}
\rho = \frac{1}{2}(I+\mathbf{r}\cdot\mathbf{\sigma})
\end{eqnarray}
where $\mathbf{r}$ is the Bloch vector of the state $\rho$. Classical states are given by
\begin{eqnarray}
\rho_c &=& p|0\rangle\langle 0|+(1-p)|1\rangle\langle 1| \\
&=& \frac{1}{2}[I +(2p-1)\sigma_z]
\end{eqnarray}
with $0\leq p\leq 1$, implying that classical states are all contained in the $z$-axis of the Bloch ball. Thus, for an arbitrary state it is enough to measure the mean values of $\sigma_x$ and $\sigma_y$ and see if they differ from zero; in this situation, the state is nonclassical. Thus, except for a null-measure set, almost all states are nonclassical. This is completely different from the notion of classicality given by $\mathsf{SU}(2)$ coherent states: for a two-level system, \textit{all} states are classical.
 
\section{Entanglement potential and controlled displacements}

\subsection{Original definition}

The term entanglement potential was first used by Asboth \textit{et al} \cite{asboth} to quantify the nonclassicality of a single-mode state by its capacity to generate entanglement when passed through a beam-splitter, while the second port is in the vaccum (or any coherent) state. The beam-splitter is a classicality preserving device and thus if the output state is entangled (nonclassical), it is mandatory that the input single-mode state displays some nonclassicality. However, some nonclassical states could be mapped into separable nonclassical states as well and there is yet no formal proof that such case is forbidden.

Also, as pointed out in \cite{asboth}, there are various differents measures of entanglement and each one will impose a different nonclassicality order for the single-system input state. Some measures can be null for some entangled states and would then assemble null nonclassicality to states that are nonclassical. Given a classicality notion, is there a measure of entanglement that corresponds consistently to the amount of nonclassicality exhibited by a single-system state? This question will remain open for the original physical context of quantum optics, but we will show that for different setups it is possible to formulate the notion of entanglement potential in a consistent way.

\subsection{Controlled displacements}

To view the notion of entanglement potential in a different perspective, we express the beam-splitter as a kind of Controlled Displacement (CD) gate. But before this, we consider the simplest case of a CD operation, the Controlled-NOT (CNOT) gate. Let us first review briefly the usual definition of a CNOT-gate, for a two-qubit system. This unitary transformation is given by $U|i,j\rangle = |i,(j \oplus_{2} i)\rangle$, where the symbol $\oplus_{d}$ means sum modulo $d$. The name comes from the property that qubits of the first subsystem determine the transformation of the second one: if the first qubit is $|0\rangle$, nothing changes in the second one, while if it is $|1\rangle$, the second qubit is flipped. We say then that the first qubit is the control qubit. The CNOT gate together with single-qubit operations are known to yield universal schemes of quantum computation \cite{nielsen}.  

Let us see the action of a CNOT gate in a general control qubit $|\psi\rangle = a|0\rangle +b|1\rangle$, with $|a|^2+|b|^2=1$:
\begin{eqnarray}
U|\psi\rangle |0\rangle &=& a|00\rangle+b|11\rangle; \\
U|\psi\rangle |1\rangle &=& a|01\rangle+b|10\rangle.
\end{eqnarray}
Thus, the output state will be entangled iff the first qubit $|\psi\rangle$ is nonclassical, when we take $\{|0\rangle, |1\rangle\}$ as our classical basis. In other words, the CNOT-gate is a classicality preserving operation. 

For a $d$-level system, the orthonormal basis $\{|0\rangle, |1\rangle, \ldots, |d-1\rangle\}$ used as reference can be given the structure of a discrete phase space \cite{vourdas}. If we define the position states as $|x\rangle$, $x=0,1,\ldots, d-1$, we can define a displacement operator $D(x')$ ($x'$ a integer value) whose action on the position states is simply to sum modulo $d$: $D(x')|x\rangle=|(x\oplus_d x')\rangle$. It is straightforward that the phase space has a toroidal structure: $D(1)|d-1\rangle = |0\rangle$. We can then express the CNOT-gate for two-qubits in terms of displacements in the following form: 
\begin{eqnarray}
U|i,j\rangle &=& |i\rangle\otimes [D_2(i)|j\rangle] \\
&=& D_1(i)D_2(i+j)|00\rangle \label{cnot}
\end{eqnarray}
where $D_i$ refers to subsystem $i$. Thus, we see that the CNOT gate is in fact a controlled displacement, where the first subsystem controls the displacement of the second one. Let us see now the situation in continuous variable's regime. When the classical basis is given by the set of coherent states, the standard classicality-preserving operation is given by the $50:50$ beam-splitter:
\begin{eqnarray*}
U_{BS}\left(\begin{array}{c}{a_1}\\{a_2}\end{array}\right) = 
\frac{1}{\sqrt{2}}\left(\begin{array}{c}
{a_2+a_1}\\
{a_2-a_1} \end{array}\right)
\end{eqnarray*}
The action of the beam splitter on the classical basis in terms of the usual displacement operators is given by
\begin{eqnarray}
U_{BS}|\alpha,\beta\rangle = U_{BS}D_1(\alpha)D_2(\beta)|00\rangle=D_1\left(\frac{\alpha-\beta}{\sqrt{2}}\right)D_2\left(\frac{\alpha+\beta}{\sqrt{2}}\right)|00\rangle \label{bs}
\end{eqnarray}
Taking $|\beta\rangle$ as the vaccum state $|0\rangle$, we have
\begin{eqnarray}
U_{BS}D_1(\alpha)D_2(0)|00\rangle = D_1\left(\frac{\alpha}{\sqrt{2}}\right)D_2\left(\frac{\alpha}{\sqrt{2}}\right)|00\rangle
\end{eqnarray}
and the beam-splitter has the structure of a CD operation. Motivated by this, we give a general definition of a CD operation in terms of its action in the displacement operators of a given classical basis:
\begin{eqnarray}
D_1(c_i)D_2(c_0) \stackrel{U_{CD}}{\rightarrow} D_1(c_i)D_2(c_i) \label{cd}
\end{eqnarray}
Without loss of generality, we will always consider the second subsystem in the state $|c_0\rangle$ of the classical basis. An advantage of the formulation (\ref{cd}) is the preffered direction of operations it imposes in a orthonormal classical basis. In this situation, a nonclassical state $|\psi\rangle=\sum_ic_i|i\rangle$ in the target qubit amounts to nothing:
\begin{eqnarray}
U_{CD}|c_i\rangle|\psi\rangle=|c_i\rangle|\psi'\rangle
\end{eqnarray}
with $|\psi'\rangle=\sum_ic_i|\sigma (i)\rangle$ and $\sigma (i)$ is a permutation of numbers $i$. This is a good feature, since we have a preffered direction of flow of operations and we can control the entanglement which is created. This will be very important for the multipartite case, as we will see shortly.

\subsection{Conversion of Superposition Number into Schmidt Number}

For an orthonormal classical basis $C=\{|0\rangle,|1\rangle,\ldots, |d-1\rangle\}$, we have defined the superposition number of a mixed state by (\ref{ns}). For a bipartite system $\mathcal{H}=\mathcal{H}_1\otimes\mathcal{H}_2$, we define the Schmidt Number of a pure state $|\psi\rangle\in\mathcal{H}$ as the rank of the reduced matrices $\rho_1=Tr_2(|\psi\rangle\langle\psi|)$, $\rho_2=Tr_1(|\psi\rangle\langle\psi|)$. An arbitrary pure bipartite state can always be written in Schmidt form $|\psi\rangle =\sum_{i=0}^rc_i|a_ib_i\rangle$, with $\{|a_m\rangle\}_{m=0}^{d_{\mathcal{H}_1}-1}$, $\{|b_n\rangle\}_{n=0}^{d_{\mathcal{H}_2}-1}$ being the respective orthonormal basis of $\mathcal{H}_1$, $\mathcal{H}_2$ and $r=min\{d_{\mathcal{H}_1}, d_{\mathcal{H}_2}\}$. Thus, the Schmidt Number of a pure state $|\psi\rangle$ is the number of non-null coefficients $c_i$ in the Schmidt decomposition of $|\psi\rangle$, or, in other words, is the Superposition Number of $|\psi\rangle$ in the Schmidt basis. The extension to mixed states is straightforward \cite{terhal2}: a mixed state $\rho$ has Schmidt number $S_N(\rho)=k$ if: (a) for any decomposition of $\rho=\sum_i p_i|\psi_i\rangle\langle\psi_i|$ at least one of the vectors $|\psi_i\rangle$ has at least Schmidt Number $k$ and (b) there exist a decomposition of $\rho$ with all vectors $|\psi_i\rangle$ with Schimdt Number at least $k$. This is equivalent to the following optimization \cite{sanpera}
\begin{eqnarray}  
S_N(\rho)=inf\{sup(r(\psi_i)):\rho=\sum_ip_i|\psi_i\rangle\langle\psi_i|\} \label{sn}
\end{eqnarray}
where the infimum is computed considering all decompositions $\rho=\sum_ip_i|\psi_i\rangle\langle\psi_i|$ and $sup(r(\psi_i))$ is the supremum value for a fixed decomposition.

Taking an arbitrary density matrix $\rho=\sum_i p_i|\psi_i\rangle\langle\psi_i|$, with $|\psi_i\rangle =\sum_j c_j^{(i)}|j\rangle$, we have that the action of the CD gate will take it to $\rho'=U_{CD}\rho U^{\dagger}_{CD}=\sum_i p_i|\Psi_i\rangle\langle\Psi_i|$, where $|\Psi_i\rangle=\sum_j c_j^{(i)}|jj\rangle$. States $|\Psi_i\rangle$ are already decomposed in Schmidt basis and their Schmidt number is simply the number of non null coefficients $c_j^{(i)}$. If we consider the subspace spanned by vectors $S=\{|kk\rangle\}_{k=0}^{d-1}$, we see that $\rho'$ is fully supported on this subspace and that its Schmidt Number is precisely its Superposition Number in the set $S$. But the Superposition Number in $S$ is simply the superposition number in the classical basis $C$. Thus, we conclude that the Schmidt number of the output $\rho'$ is precisely the superposition number of the input $\rho$. The CD gate fully converts nonclassical states into entangled ones, in a consistent way. This results depends on the orthonormality of the classical basis. For continuous variables, we see that the conversion is not one-to-one. If we take a Fock state $|n\rangle$, which has an infinite superposition number in the classical basis of coherent states, we get as output of a $50:50$ beam-splitter 
\begin{eqnarray}
|\phi_n\rangle=U_{CD}|n\rangle=\sum_{k=0}^n \sqrt{{n\choose k}}|k,n-k\rangle
\end{eqnarray}
which clearly has finite Schmidt rank. 

As a corollary, we get that for an orthonormal classical basis, if we detect the nonclassicality of a state $\rho$ via an observable $M$, i.e., $\lambda(M)<Tr(M\rho)$, we have that state $\rho'=U_{CD}\rho U^{\dagger}_{CD}$ will be surelly entangled. Thus, we can use a local observable to detect the entanglement of the output state $\rho'$, which represents an enormous practical simplification. Also, even when an orthonormal classical basis or the corresponding CD operation are not available, one can see some entangled states as the output of a virtual CD gate acting in a virtual nonclassical state and their entanglement can be detected in a simpler manner.

\section{Multipartite entanglement}

Previous considerations can be further extended to multipartite systems, with good simplifications and possibilities of classification. For multipartite systems it is not possible in general to have a Schmidt decomposition. Let us see first for simplicity the case of three qubits. In this case, there are two classes of tripartite entangled states, the GHZ class
\begin{eqnarray}
GHZ = \frac{|000\rangle +|111\rangle}{\sqrt{2}}
\end{eqnarray}
and the $W$ class:
\begin{eqnarray}
W=\frac{|100\rangle+|010\rangle+|001\rangle}{\sqrt{3}}
\end{eqnarray}
These classes are inequivalent under SLOCC operations \cite{dur}. We will see how to create states in each class using local nonclassical operations and CD operations.

Pure states representatives of each class can be generated through right use of single-system's nonclassicality and concatenated CNOT gates. Starting with a state $|\psi\rangle = (a|0\rangle +b|1\rangle)|00\rangle$ we have
\begin{eqnarray}
U^{2,3}_{CNOT}U^{1,2}_{CNOT}|\psi\rangle = U^{2,3}_{CNOT}(a|00\rangle+b|11\rangle)|0\rangle = a|000\rangle + b|111\rangle
\end{eqnarray}
which is a state in the GHZ class. The class of W states is reached in a more involved way. We first apply a CNOT to the first and second qubits:
\begin{eqnarray}
|\psi'\rangle = U^{1,2}_{CNOT}|\psi\rangle = (a|00\rangle+b|11\rangle)|0\rangle
\end{eqnarray}
Then we apply the following local invertible nonunitary operation to the second qubit:
\begin{eqnarray}
T=\left(\begin{array}{c c}
{1}&{0}\\ 
{1}&{1}\end{array}\right) \label{T}
\end{eqnarray}
and we are left with the following state
\begin{eqnarray}
|\psi''\rangle = I\otimes T\otimes I|\psi'\rangle = [a|0\rangle (|0\rangle+|1\rangle)+b|11\rangle]|0\rangle
\end{eqnarray}
We now apply the CNOT gate to the second and third qubits:
\begin{eqnarray}
|\psi'''\rangle = U^{2,3}_{CNOT}|\psi''\rangle = a|0\rangle (|00\rangle +|11\rangle)+b|111\rangle
\end{eqnarray}
Finally, we apply the CNOT to the first and third qubits:
\begin{eqnarray}
U^{1,3}_{CNOT}|\psi'''\rangle = a|0\rangle(|00\rangle +|11\rangle)+b|1\rangle|10\rangle
\end{eqnarray}
This last operation comes from the fact that the W state is pairwise entangled, so we need to entangle the first and third qubits as well. This simple example separates different classes of entanglement, considering the different options of classicality definition that come to play during the creation of the state. While the GHZ state is obtained simply distributing the nonclassicality of a single state through the other subsystems, the W state demands a nonunitary local change of basis along the way and we can interpret this as a different notion of nonclassicality along the way. 

The transformation $T$ in (\ref{T}) seems arbitrary at first, but there is a nice way to account for it. To obtain from the classical basis $\{|0\rangle, |1\rangle\}$ another vectorial basis, one needs to apply an invertible operation to each element. Any linear transformation admits a Jordan normal form $J=\bigoplus_i J^{(d_i)}_i$, where $J_i^{(d_i)}$ are $d_i\times d_i$ ($\sum_i d_i=d_{\mathcal{H}}$) Jordan blocks given by
\begin{eqnarray}
J_i^{(d_i)} = \left(\begin{array}{c c c c}
{\lambda_i}&{}&{}&{}\\
{1}&{\lambda_i}&{}&{}\\
{}&{1}&{\lambda_i}&{}\\
{}&{}&{\ddots}&{\ddots}
\end{array}\right)
\end{eqnarray} 
We see that for a two-level system, there are two possibilities:
\begin{eqnarray}
T_1 = \left(\begin{array}{c c}
{\lambda_1}&{0}\\ 
{1}&{\lambda_1}\end{array}\right); \ \ \left(\begin{array}{c c}
{\lambda_1}&{0}\\ 
{0}&{\lambda_2}\end{array}\right)
\end{eqnarray}
each corresponding to one class of entanglement. This is in close relation to the classification in terms of sub-Schmidt rank given in \cite{cornelio}. We proceed in order to generate all the classes of tripartite entanglement of two qutrits and a qubit. A three-level system has the following possibilities of Jordan normal forms:
\begin{eqnarray*}
T_1 = \left(\begin{array}{c c c}
{\lambda_1}&{0}&{0} \\
{1}&{\lambda_1}&{0} \\
{0}&{1}&{\lambda_1}\end{array}\right); \ \ T_2 = \left(\begin{array}{c c c}
{\lambda_1}&{0}&{0} \\
{0}&{\lambda_2}&{0} \\
{0}&{1}&{\lambda_2}
\end{array}\right); \ \ T_3 = \left(\begin{array}{c c c}
{\lambda_1}&{0}&{0} \\
{0}&{\lambda_2}&{0} \\
{0}&{0}&{\lambda_3}
\end{array}\right)
\end{eqnarray*}
Since the ordering of systems matter, we first consider the $3\otimes 3\otimes 2$ case. Starting with a state of the form $|\psi\rangle = a|0\rangle + b|1\rangle +c|2\rangle$, we apply the CD gate to the first and second qutrits:
\begin{eqnarray}
U_{CD}^{1,2}|\psi\rangle |00\rangle= (a|00\rangle +b|11\rangle+c|22\rangle)|0\rangle \label{inicial1}
\end{eqnarray}
Applying, in this order \footnote{For simplicity, we set $\lambda_i=1$ and do not care about normalization.}, $I\otimes T_1\otimes I$, $U^{2,3}_{CD}$, $U^{1,3}_{CD}$, we obtain
\begin{eqnarray}  
(a|00\rangle +b|11\rangle +c|22\rangle)|0\rangle + (a|01\rangle +b|12\rangle)|1\rangle
\end{eqnarray}
If we apply to (\ref{inicial1}) the operations $I\otimes T_2\otimes I$, $U^{2,3}_{CD}$, $U^{1,3}_{CD}$, we get
\begin{eqnarray}
(a|00\rangle +b|11\rangle +c|22\rangle)|0\rangle + b|12\rangle |1\rangle
\end{eqnarray}
Now we consider the $2\otimes 3\otimes 3$ situation. Starting with a state $|\psi\rangle = a|0\rangle +b|1\rangle$, we apply the $U_{CD}^{1,2}$ gate:
\begin{eqnarray}
U_{CD}^{1,2}|\psi\rangle |00\rangle = a|00\rangle +b|11\rangle \label{inicial2}
\end{eqnarray}
Applying, in this order, $I\otimes T_1\otimes I$, $U^{2,3}_{CD}$, $U^{1,3}_{CD}$, we obtain
\begin{eqnarray}
a|0\rangle(|00\rangle)+b|1\rangle(|12\rangle +|20\rangle)
\end{eqnarray}
while applying $I\otimes T_2\otimes I$, $U^{2,3}_{CD}$, $U^{1,3}_{CD}$ to (\ref{inicial2}), the result is
\begin{eqnarray}
a|000\rangle +b|1\rangle(|11\rangle +|22\rangle)
\end{eqnarray}
The case $3\otimes 2\otimes 3$ does not produce different results. The four states obtained above corresponds precisely to the four inequivalent classes of tripartite entangled states for two qutrits and a qubit \cite{cornelio}.  
\section{Conclusion}

In this work, we showed the existence and constructed nonclassicality witnesses for a given notion of nonclassicality of a system. These witnesses are constructed from the observables available in a given situation, having thus a great practical appeal. Some examples of usual nonclassical states were detected by the method in a simple way. We stress that from the observables used many more conclusions could be drawn. For example, a pure nonclassical state $|\psi\rangle$ were detected easilly defining its rank one projector $M_{\psi}=|\psi\rangle\langle\psi|$. For a nonclassical $|\psi\rangle$, it is trivial that $\lambda(M_{\psi})<1=Tr(M_{\psi}|\psi\rangle\langle\psi|)$. However, with $M_{\psi}$ we can detect many mixed states. In finite dimensional systems, if we define
\begin{eqnarray}
\rho=p\frac{I}{d}+(1-p)|\psi\rangle\langle\psi|
\end{eqnarray}
we have that $Tr(M_{\psi}\rho)=pd^{-1}+(1-p)$. So, whenever 
\begin{eqnarray}
p<\frac{1-\lambda(M_{\psi})}{1-1/d}
\end{eqnarray}
we will have a nonclassical state. Another example would be the following state
\begin{eqnarray}
\rho=p\rho^{\bot}+(1-p)|\psi\rangle\langle\psi|
\end{eqnarray}
with $\rho^{\bot}$ an state orthogonal to $|\psi\rangle$. Then the condition of nonclassicality is $p<1-\lambda(M_{\psi})$.

In the discussion concerning CD operations, we fixed the second port in a classical basis element. Doing this, the notion of entanglement potential was consistently extended. However, for other applications one could think about more general states in the second port. One simple remark that should be considered is that whenever there is an eigenvector of the generator of displacement operator in the second port of a CD gate, the output will be a separable state. These states are, in the language of \cite{vourdas}, the momentum states of the discrete phase-space, forming a mutually unbiased basis with respect to the classical basis.  

For multipartite states, we showed how to construct the different classes of tripartite states by applying different nonclassical operations to the classical basis of the second subsystem and using CD gates. We will let for future works to determine if this approach exhausts all the classes of multipartite entanglement. One should consider, however, that this approach has a practical appeal considering the diverse classes of multipartite entanglement that can be created in practice, given that one can perform some restricted set of nonclassical operations. For example, in quantum optics it is fair to say that squeezed states and Fock states are easy to produce in laboratory. The situation is not the same for cat states. Thus, the experimentalist is somewhat restricted to the entanglement classes produced by combinations of squeezing and photon-addition.

\section{Acknowledgments}

I would like to express my gratitude to T. F. Viscondi, M. A. Marchiori and M. F. Cornelio for insights and discussions about the topics of this manuscript. This work was financed by brazilian agency CAPES.

\end{document}